# Depressed surface magnetic properties in $La_{2/3}Ca_{1/3}MnO_3$ thin films probed by X-ray magnetic circular dichroism in reflection.

S. Valencia*†, A. Gaupp and W. Gudat

BESSY, Albert-Einstein-Str. 15, 12489, Berlin, Germany.

Ll. Abad, Ll. Balcells and B. Martínez

Institut de Ciència de Materials de Barcelona, ICMAB – CSIC,

Campus UAB, 08193 Bellaterra, Spain

**Abstract**

Surface magnetic properties of perovskite manganites have been a recurrent topic during last years since they play a major role in the implementation of magnetoelectronic devices. Magneto-optical techniques, such as X-ray magnetic circular dichroism, turn out to be a very efficient tool to study surface magnetism due to their sensitivity to magnetic and chemical variations across the sample depth. Nevertheless, the application of the sum rules for the determination of the spin magnetic moment might lead to uncertainties as large as 40% in case of Mn ions. To overcome this problem we present an alternative approach consisting of using X-ray magnetic circular dichroism in reflection geometry. A fit of the data by using a computer code based in a 4X4 matrix formalism shows that surface and interface roughness are of major relevance for a proper description of the experimental data and a correct interpretation of the results. By using such an approach we demonstrate the presence of a narrow surface region with strongly depressed magnetic properties in $La_{2/3}Ca_{1/3}MnO_3$ thin films.

* Corresponding author: sergio.valencia@gmx.de
† Actual address: *Institut für Experimentalphysik, Freie Universität Berlin, Arnimallee 14, 14195 Berlin, Germany*

## Introduction

Rare-earth manganese perovskites, also known as manganites, are thought to be a real alternative for substituting and improving magnetic systems based on the giant magnetoresitance effect. Their half metallic character [1-3], i.e. 100% spin polarization make them very appealing materials to be implemented in the form of thin films as spin devices, e.g. as magnetic tunnel junctions. In such systems the electrical resistance across the insulator barrier can strongly be modified at low temperature by applying magnetic fields of only a few Oe. It is theoretically expected that the effect persists whenever the manganites are ferromagnetic and decreases when their Curie temperature is approached. However, it is experimentally found that it already vanishes far below $T_C$ [3, 4] delaying the implementation of manganite systems in spintronic applications.

The implementation of magnetoelectronic devices based on these mixed valence oxide materials requires in most of the cases the use of thin films. Currently one of the most studied manganite compounds is $La_{2/3}Ca_{1/3}MnO_3$ (LCMO). This compound presenting a mixed $Mn^{3+/4+}$ valence is ferromagnetic below 270 K in case of bulk material. Nevertheless, its magnetic transition temperature ($T_C$) is strongly reduced when grown in the form of thin films [5-7]. In order to achieve a successful implementation of LCMO thin films in spintronic devices a good knowledge and control of the surface and interfaces of these materials is necessary. This is especially relevant in case of tunnel magnetic junctions where the magnetotransport properties of the macroscopic system strongly depend on the magnetic and chemical homogeneity of the various interfaces. To this respect Calderón et al. have theoretically demonstrated that a weakening of the double exchange mechanism concomitant with an increase of the antiferromagnetic Heisenberg coupling might take place at the surface of manganites if the braking of the crystal symmetry results in a localized oxygen deficiency for the outermost layer [8], i.e. the apical oxygen of the $MnO_6$ octahedron is missing. This induces a splitting of the Mn 3d $e_g$ sub-levels lowering the energy corresponding to the $3d_{3z2-r2}$ orbital which may trap part of the charge. The breaking of the symmetry thus might lead to a reduction of the surface magnetization concomitant with an increase of the surface resistance.

Currently the most widespread magneto-optical method for surface magnetic characterization is X-ray magnetic circular dichroism (XMCD) in absorption geometry [9]. XMCD is the difference of absorption of left and right circularly polarized radiation which takes place in materials magnetized parallel to the propagation direction of the radiation. As demonstrated formally by Brouder and Kappler [10], the same effect is obtained when fixing the helicity of the incoming photons but changing the sign of the magnetization. Within the soft X-ray region (ca. 200 eV-2000 eV) the $L_3$ and $L_2$ electronic transitions ($2p_{3/2}\rightarrow 3d$ and $2p_{1/2}\rightarrow 3d$, respectively) of transition metals (TM) are accessible. Here the excited 2p electrons directly probe the partially occupied 3d states. Therefore XMCD at the Mn L-edges (ca. 642 eV and 655 eV, respectively) becomes an ideal tool for an element selective magnetic characterization of manganite compounds since the Mn 3d levels are at the origin of their magnetic properties. The development of the so called sum rules by Thole et al. [11] and Carra et al. [12] is the ground to understand why this technique has become so popular. Such rules allow to obtain quantitative magnetic information from the XMCD. Moreover, the sum rules allow to separate the spin and orbital contributions to the magnetization, $m_{spin}$ and $m_{orb}$, respectively.

In spite of such impressive capabilities which includes the almost errorless determination of $m_{orb}$ by using the sum rules, , XMCD is not free of uncertainties. E.g. the determination of $m_{spin}$ can be affected by an uncertainty as large as 30% for the specific case of Mn due to the relatively small "spin-orbit" splitting of the $L_3$ and $L_2$ edges [13, 14]. Moreover the $m_{spin}$ sum rule requires the magnetic dipole operator $\langle T_Z \rangle$ which measures the sphericity of the spin distribution around the TM ion. This term is usually neglected invoking symmetry or temperature arguments [15, 16]. Nevertheless, a large contribution of $<T_Z>$ to $m_{spin}$ is expected in case of surface sensitive measurements, such as those performed by recording the yield current (total electron yield, TEY), or in case of low temperature experiments [16]. Precisely these are the experimental conditions which are otherwise required for a fully surface magnetic characterization of mixed valence La-Mn manganites such as LCMO. For the case of $Mn^{4+}$ it is theoretically expected that within the above experimental conditions, i.e. surface sensitive measurement at low temperature, ignoring the $<T_Z>$ operator leads to an error in the stimation of $m_{spin}$ as large as 40% [16]. Not only the experimental determination of $<T_Z>$ is a very complicate task but also ab-initio electronic structure calculations of $<Tz>$ since knowledge of the energy band distribution at the surface/interface are required.

Magneto-optical techniques in reflection geometry within the soft X-ray region lead to magneto-optical effects larger than usually observed in absorption experiments [17]. This provides us with a high magnetic sensitivity as demonstrated for example by the observation of an induced magnetic moment in Carbon in C/Fe multilayers [18] or more recently by the detection of magnetically induced signatures at the O, La and Ca ions of LCMO [19]. Unfortunately, there is no equivalent to the sum-rules for reflection-XMCD (R-XMCD), i.e. no quantitative information can be extracted. The reason is that in reflection geometry, as opposed to absorption, the shape of the XMCD spectrum depends not only on absorption but also on dispersion. The roughness of the various interfaces and interference effects also play a major role. Consequently, a modelling of the data to obtain physical information is necessary. To this respect Zak et al. developed a matrix formalism suitable for simulation of magneto-optical effects in reflection and transmission geometries for incident polarized light and arbitrary direction of the magnetization. Computer codes based on that formalism has been used to demonstrate interference effects in longitudinal magneto-optical Kerr effect spectrum (L-MOKE) [20], the close relation between L-MOKE ellipticity and rotation spectra with R-XMCD and transversal-MOKE (T-MOKE) [21] and more recently for demonstrating the existence of surface magnetic dead layers in case of the layered manganite $La_{2-2x}Sr_{1+2x}Mn_2O_7$ [22]. This later case deserves special attention. A change in the sign of the R-XMCD asymmetry, defined as AR-XMCD=$(R^+-R^-)/(R^++R^-)$ with $R^+$ ($R^-$) being the reflectance for magnetization parallel (antiparallel) to the propagation direction of the incoming radiation, was observed when the angle of incidence was changed from 5° to 16°. This change of sign was interpreted as originating from interference effects between a non-magnetic surface layer 1 nm thick and the homogeneous magnetization of the bulk of the sample. Although this interpretation might be correct in case of layered manganites which can easily be cleaved thus presenting an atomically smooth surface, we will show in the following that interface roughness can also lead to effects very akin to those produced by magnetic dead layers, i.e. a change of sign in the R-XMCD asymmetry.

In this work we present an alternative approach for surface element-selective investigation of magnetic properties avoiding intrinsic absorption-related problems. Measurement of XMCD spectra in reflection geometry and subsequently fit of the experimental data are used for analyzing surface magnetization in LCMO manganite thin films grown on various substrates. The fit is done by using a computer code based on Zak´s formalism [23] for the description of magneto-optical effects. Within the computer code, and as a main difference with respect to previous existing codes we explicitly consider the roughness of the various interfaces existing in the samples. The fits, irrespective to the detailed properties of the LCMO sample, point to the presence of a surface layer with reduced magnetization with respect to the bulk of the sample. These results are validated by comparing experimental T-MOKE spectra with simulated ones using the fit parameters obtained from R-XMCD data as input. This surface layer with strongly degraded magnetic properties might be at the origin of the observed limited functionality of magnetic devices based on manganese perovskites, such as tunnel junctions or spin filters [4].

**Experimental:**

The LCMO samples (see table 1 for details) were grown by means of rf magnetron sputtering on top of three different single crystalline substrates, $SrTiO_3$ (001)-oriented (STO), $LaAlO_3$ (001)-oriented (LAO) and $NdGaO_3$ (110)-oriented (NGO). Due to the lattice mismatch between the manganite ($a_{LCMO}$= 0.386 nm) and the substrates ($a_{STO}$= 0.3905 nm, $a_{LAO}$= 0.379 nm and $a_{NGO}$= 0.386 nm) the material grows under in-plane biaxially tensile strain, compressed or almost strain free, respectively. The thickness of the LCMO layer ranges from 6 to 50 nm.

During deposition the substrate temperature was kept at 800 °C. The pressure of the sputter gas was 330 mTorr (Ar- 20% $O_2$). Subsequently, films were *in situ* annealed at 800 °C at an oxygen pressure of 350 Torr for 1 h. Afterwards, they were cooled down to room temperature at a rate of 15 °C/min at the same oxygen partial pressure. One LCMO/STO film of t=50 nm and the t=16 nm LCMO/LAO film were in addition annealed in air for 2 h at $T_{ann}$=1000 °C, with heating and cooling ramps of 5 °C/min which is known to improve the magnetic and transport compared to as-grown films [7].

The thickness of the thin film samples and their surface roughness were deduced from grazing incidence Cu $K_\alpha$ x-ray reflectometry (XRR). The out-of-plane cell parameter *c* was determined from x-ray diffraction experiments using the LCMO (004) reflection. The in-plane cell parameter *a* was deduced from reciprocal space maps (Q-plots) around the (103) manganite reflection. Magnetization curves, *M*(*H*) at *T*=10 K and *M*(*T*) with an applied field of *H*=5000 Oe, were measured by using a superconducting quantum interference device magnetometer (Quantum Design).

The synchrotron radiation experiments were performed at low temperature (T ≈80 K) at the dipole beamline PM3 of BESSY. The spectral resolution at the Mn 2p edges was roughly E/ΔE=2500. The degree of polarization was $P_{Circ}$=0.87±0.03 (right helicity) for the R-XMCD measurements and $P_{Lin}$>0.99 (p-geometry) for the T-MOKE measurements. The radiation impinged upon the sample at a fixed grazing incidence angle θ=13°. Two coils allowed magnetic saturation of the sample parallel to its surface, either parallel (longitudinal) or perpendicular (transverse) to the plane of incidence in case of R-XMCD or T-MOKE experiments. We used the BESSY ultrahigh vacuum polarimeter chamber [24] which allows the simultaneous measurement of reflectivity and of absorption by using total electron yield (TEY) detection mode. For TEY the photoexcited drainage current of the sample was recorded while the sample was kept at a potential of −95 V with respect to the chamber. In order to avoid interference of the magnetic fields with the electrons collected in TEY mode, the spectra were obtained in magnetic remanence.

The R-XMCD and T-MOKE spectra have been fit and simulated, respectively, by using the computer code mentioned above. The Minuit routine developed at CERN [25] has been used as the fitting algorithm. The optical and magneto-optical constants (index of refraction) for each of the materials in the sample are required as input data for the computer code. The non-magnetic and magnetic absorptive parts (β and Δβ, respectively) of the index of refraction for the LCMO layer have been obtained for each of the samples from their own TEY spectra after proper normalization of the pre- and post-edge regions to data from the Henke table [26]. The dispersive part (δ and Δδ) has been computed from a Kramers-Kronig transformation of the absorption spectra [27]. The optical constants for the substrates have also been obtained from [26].

**The computer code. Roughness effects:**

The 4X4 matrix formalism introduced by Zak et al. allows the simulation of magneto-optical effects with arbitrary direction of the magnetization and polarization of the incoming radiation. It is based on the conservation of the tangential components of the electric and magnetic fields of the radiation at the interface between two different media, e.g. *w* and *w+1* [23]. This is described by the so-called medium boundary matrix $\boldsymbol{A_w}$. The absorption and dispersion of the radiation within a given layer *w* is accounted by another matrix $\boldsymbol{D_w}$ called propagation matrix. In case of a multilayer system with *nl* layers (*w*=1,2…nl) the incoming radiation has its origin in an incident medium *i*, e.g. vacuum, propagates through the *nl* layers (from *w*=1 to *w*=*nl*) and ends up in the final medium *f*, e.g. the substrate. The knowledge of the $\boldsymbol{A_w}$ and $\boldsymbol{D_w}$ matrices for each of the interfaces and layers allows to deduce the polarization state of the radiation at the final medium ($P_f$) if the initial state ($P_i$) is known:

$$P_i = A_i^{-1} \prod_{w=1}^{nl} \left( A_w D_w A_w^{-1} \right) A_f P_f = M P_f \qquad with \qquad M = \begin{pmatrix} G & H \\ I & J \end{pmatrix} \tag{1}$$

being $\boldsymbol{G}$, $\boldsymbol{H}$, $\boldsymbol{I}$ and $\boldsymbol{J}$ 2x2 matrices.

The reflection coefficients for the entire multilayer system are obtained by:

$$\begin{pmatrix} r_{ss}^{nl} & r_{sp}^{nl} \\ r_{ps}^{nl} & r_{pp}^{nl} \end{pmatrix} = I \cdot G^{-1} \tag{2}$$

being $r^{nl}_{ij}$ the reflection amplitude coefficient for incident *i* (*s* or *p*) and reflected *j* (*s* or *p*) polarized light for the entire multilayer system.

Within this formalism, the roughness of the different layers is not considered. Nevertheless, in scattering experiments roughness effects might be important since they contribute to the dispersion of the incoming radiation. In such a case both, the transmitted and the specular reflected radiation decrease [28] affecting as we will show magneto-optical effects.

Classically, i.e. within the Parrat formalism, the roughness of a given layer *w* is accounted by multiplying its reflection amplitude coefficient $r^w$ by a roughness modelling factor $\chi^w$ which decreases the amplitude of the reflected wave. The reflection coefficient for the whole system is then calculated using the recursive formulae [29]:

$$R^w = \frac{R^{w-1} + r^w \cdot \chi^w}{1 + R^{w-1} + r^w \cdot \chi^w} \qquad (3)$$

where $R^w$ is the total reflection amplitude coefficient for the system including layers 1,2,…*w.*

To mimic this procedure within Zak´s formulation is necessary to rewrite eq. (1) in order to obtain the individual reflection amplitude coefficients $r^w$ for each of the layers as:

$$P_i = \prod_{w=1}^{nl} M^w P_f = MP_f \qquad with \qquad M^w = D^w \left(A^W\right)^{-1} A^{w+1} \qquad (4)$$

where each of the $\boldsymbol{M^w}$ matrices correspond to a given medium *w* including the interface with medium *w+1*. As in the case of the whole layer system the $\boldsymbol{M^w}$ matrices can be rewritten as a set of elements of 2X2 matrices, namely $G^w_{ij}$, $H^w_{ij}$, $I^w_{ij}$, and $J^w_{ij}$ (with i=j=1,2), which allows the calculation of the individual reflection coefficients by $\boldsymbol{r^w=I^w(G^w)^{-1}}$.

In case of unpolarized incoming radiation ($r^w_{pp}= r^w_{ss}=r^w$) on a single non-magnetic layer ($r^w_{sp}=r^w_{ps}=0$ and off-diagonal elements of the 2X2 matrices ***G***, ***H***, ***I***, and ***J*** are all zero), the explicit analytical form of the reflection coefficient system within Zak's formulation leads to:

$$R^1 = r^1 = \frac{I^1_{ii}}{G^1_{ii}} = \frac{H^1_{ii}}{J^1_{ii}} \qquad (5)$$

This expression holds for any individual layer thus the reflection amplitude coefficient of a layer *w* within a *nl* multilayer system reads $r^w = I^w_{ii} / G^w_{ii} = H^w_{ii} / J^w_{ii}$ .

The inclusion of roughness effects for a given layer *w* within Zak's formulation is then achieved by multiplying the $I^w_{ii}$ and $H^w_{ii}$ terms of the individual $\boldsymbol{M^w}$ matrix by the roughness factor $X^w$. Thereafter the product of the modified $\boldsymbol{M^w}$ matrices is recalculated according to eq. 4 allowing to obtain a reflection coefficient for the whole system including roughness effects by using expression 2. The validity of the approach has been checked by comparison of simulated spectra within Zak's formalism including the roughness with those simulated using a computer code based on Parrat's formalism. The agreement is perfect in angular and energy spectra. Within the computer code, the roughness factor $X^w$ can be calculated according to two different models, i.e. Debye-Waller or Névot-Croce [28].

We have investigated the effect of interface roughness in the R-XMCD asymmetry spectrum at the Mn L-edge of LCMO. Figure 1 shows various computer simulations for an LCMO 16 nm layer on a LAO substrate. The optical and magneto-optical data used as input parameters [23] have been obtained from TEY experiments. We have observed that increasing film/ substrate interface roughness ($\sigma_{f/s}$) from 0 nm to 4 nm produces an increase of the magnitude of the magneto-optical effect without affecting its shape. On the other hand, a similar increase of the surface roughness ($\sigma_{Surface}$) strongly modifies the shape of the R-XMCD asymmetry curve. In particular figure 1 demonstrates that $\sigma_{Surface} > 1$ nm can change the sign of the asymmetry curve at certain energy regions mimicking the effect caused by the presence of a non-magnetic surface layer in atomically flat films [22]. This is demonstrated by the simulations depicted in figure 2. We present R-XMCD asymmetry spectra simulations for a 6 nm and a 50 nm LCMO layers on a STO substrate. Large spectral differences are observed when comparing the curves corresponding to an homogenous magnetic film to another of identical thickness but with topmost non magnetic layer of 1 nm. These results demonstrate both the strong sensitivity of R-XMCD to magnetic inhomogeneities and the clear requirement of taking into consideration roughness effects explicitly in order to investigate the existence of poorly magnetic surface layers in thin films. In the following section we will use the above approach to demonstrate the existence of a topmost surface layer with strongly degraded magnetic properties in LCMO thin films.

**Results and discussion:**

In Table 1 we summarize the magnetic and structural properties of the investigated LCMO films. Films grown on NGO (LCMO/NGO) -zero lattice mismatch- and those annealed in air at high temperature ($T_{ann}$) exhibit bulk-like magnetic properties. On the contrary, as-grown films on STO (LCMO/STO) substrates exhibit thickness-dependent depressed magnetic properties as previously reported [5-7].

All magneto-optical experiments have been performed at T=80 K, thus deep in the ferromagnetic regime of all films. Magnetic hysteresis loops for all films have been obtained by means of R-XMCD and T-MOKE at different photon energies across the Mn L-edge, i.e. ≈640 eV, 642 eV and 655 eV (see inset of figure 3). According to the magnitude of the absorption β the penetration depth at these energies $P_{depth}=(\lambda\sin\theta)/(4\pi\beta)$ is calculated to be 20 nm, 8 nm, and 13 nm, respectively. Figure 3 shows as an example the hysteresis obtained by means of R-XMCD for the 16 nm thick LCMO/LAO annealed sample. For all three energies the normalized hysteresis loops look similar. No changes on the shape of magnetization curve neither in the coercive nor in the saturation field values are observed. Similar results have also been obtained for all the other samples. Due to the relatively large penetration depth of the X-ray radiation in comparison with the thickness of the samples, these results indicate a similar magnetic domain rotation for all the ferromagnetic regions of the film. We note that the data depicted in figure 3 do not exclude the presence of antiferromagnetic or non-magnetic regions.

For comparison purposes, table 2 shows the value of $m_{orb}$ and $m_{spin}$ obtained by means of XMCD in the TEY detection mode in absorption geometry by using the sum rules assuming $<T_Z>=0$. As expected from the less than half filling of the 3d band the orbital and spin magnetic moments are antiferromagnetically aligned as evidenced by the difference in sign between $m_{orb}$ and $m_{spin}$. The orbital magnetic moments, although small, are non-zero for all the samples. The non-complete quenching of the angular momentum because of the Jahn-Teller distortion of the $MnO_6$ octahedron leads to this small contribution. The values reported here are in agreement with those previously reported by Song et al. [30] and Koide et al. [15].

As seen in Table 2 the total magnetic moment $m_{xmcd}=m_{spin}+m_{orb}$ obtained by means of XMCD using TEY and thus, in a surface sensitive way, is well below the averaged macroscopic magnetization measured by means of SQUID magnetometry for all films. Nevertheless, this is not a conclusive proof of the existence of a surface region with reduced magnetic properties as compared to the rest of the film. As commented, the uncertainties introduced in the calculation of $m_{spin}$ and thus in $m_{xmcd}$ by ignoring the $<T_Z>$ term are too large to allow a definitive conclusion. As an alternative we have fit the XMCD spectra in reflection geometry obtained simultaneously with that on absorption for each of the samples by using the computer code described above. Two possibilities have been explored namely; 1) a single homogeneous magnetic LCMO layer and 2) a film composed of two manganite layers with different magnetic properties. As fitting variables for the model we have used the total thickness of the film (t), the roughness at the film/substrate interface ($\sigma_{f/s}$), the roughness at the surface ($\sigma_{surface}$), the thickness of the topmost layer $t_2$, the roughness of the

layer(1)/layer(2) interface ($\sigma_{1/2}$) and a multiplicative factor for the magneto-optical constants ($\Delta\beta$ and $\Delta\delta$) of both layers ($MO_1$ and $MO_2$, respectively) accounting for possible different magnetization properties.

The fit minimizes $\chi^2$ defined as $\chi^2_i = [(AR-XMCD_{(exp)} - AR-XMCD_{(theor)}) / AR-XMCD_{(exp)}]^2$, being $AR-XMCD_{(exp)}$ and $AR-XMCD_{(theor)}$ the experimentally measured asymmetry and that obtained by means of the fitting procedure, respectively. The subindex i=1,2 stands for the models described above. The upper panels of figures 4 and 5 depict the fit results together with the experimental AR−XMCD spectra; very good agreement is observed in all cases. Table 3 summarizes the results of the fit. In order to compare on equal footing the minimized $\chi^2_i$ functions they have been normalized to the number of degrees of freedom, i.e. *number of points - number of fitting parameters - 1*. The lower values for the minimization function $\chi^2_i$ and hence the best fit results have systematically been obtained for the second fitting scenario, i.e. a film composed of two LCMO layers ($\chi^2_2$) with magneto-optical constants and thus, magnetization saturation differing by a factor MO(2)/MO(1). We note that fit values for film thickness as well as thicknesses corresponding to the roughness are in agreement with those obtained by means of XRR (see table 1), validating the here used modelling for the roughness of the various interfaces.

The results presented in Table 3 reveal the presence in all LCMO films of a thin surface layer ranging from 0.5 nm to 2 nm with depressed magnetic properties as compared with that of the rest of the film, i.e. MO(2)/MO(1)<1. The degradation of the surface magnetization is more severe in case of the as-grown films on STO. In this case the surface layer is at all not magnetic, i.e. MO(2)/MO(1)=0 that is a magnetic dead layer.

Although it is well known that high temperature annealing of as-grown films as well as growth on NGO substrates (zero-lattice mismatch) approach the macroscopic magnetic properties towards those of the bulk (see table 1) [7], the results of the fit reveal that even in these cases a thin surface region presents depressed magnetic properties. We note however that MO(1)/MO(2)≠0 thus, revelling a degraded surface magnetization as opposite to the magnetic dead layer observed for as-grown films on STO.

The presence of a surface region with reduced magnetization as compared to the rest of thee film, independently of the substrate being used, the strain conditions or the post-annealing treatment points to an intrinsic property of LCMO layers. This might be explained by the breaking of the crystal symmetry at the surface. Calderón et al. showed theoretically that a surface oxygen deficiency caused by the breaking of the crystal symmetry might lead to a weakening of the double exchange ferromagnetic interaction [8]. Concomitant with it the antiferromagnetic Heisenberg coupling would be reinforced leading to the observed decrease of magnetization at the surface. It was also shown that the affected region would extend from 1 to 3 unit cells from the surface, that is between ca. 0.4 nm to 1.2 nm. These results are thus in clear agreement with our finding.

In order to support our fit results we have simulated asymmetry T-MOKE spectra (AT-MOKE) for all samples using the fit parameters summarized in table 3 and compared with experimental ones. Results are shown in the bottom panels of figures 4 and 5. The overall agreement between both spectra is very good thus, confirming the presence of a surface layer with reduced magnetization deduced from R-XMCD experiments.

As pointed out by Calderon et al [8]. The presence of a magnetically poor surface layer must also induce a higher resistance state within this region. The increased strength of antiferromagnetic interactions might lead to partial localization of the mobile charge because an splitting of the Mn 3*d* $e_g$ sub-levels lowering the energy corresponding to the $3d_{3z2-r2}$ orbital. Indeed this surface can even behave as an electrical insulator. To this respect, Abad et al. have recently measured the surface resistance of LCMO thin films grown on LAO substrates [30, 31] using the very same deposition conditions as in this work. The surface resistance of the manganite layer was measured by using an atomic force microscope operated in current sensing mode. The measured current/voltage curves showed clear indication of a tunnelling-like transport. The thickness of the insulator tunnel barrier $t_{tunnel}$ was determined to be 3.3 nm in case of as-grown films and $t_{tunnel}$=0.5 nm for annealed samples ($T_{ann}$=1000 °C). Unfortunately, we do not have data for as-grown LCMO films on such a substrate to compare with the electric transport results, because the magnetic field necessary for magnetic saturation exceeds the capabilities of the polarimeter chamber. Nevertheless, all as-grown samples studied here (on STO and NGO substrates) show a surface of reduced magnetization with thickness ranging from 1 to 2 nm which roughly agrees with the values reported by Abad et al. for as-grown films. Moreover, our annealed 16 nm LCMO/LAO sample shows indeed a poor magnetic layer of 0.5 nm in clear agreement with Abad et al. results.

**Conclusions:**

It is a fact that the use of the sum rules for XMCD data in absorption geometry obtained by means of surface sensitive techniques, such as TEY, fails to obtain reliable values for the surface spin magnetization. The breaking of the crystal symmetry at the surface implies non-vanishing values for the magnetic dipole operator, $<T_Z>$, which is usually ignored. The related uncertainties preclude reliable estimations of the surface magnetization deficit in manganite systems. Here it has been shown that magneto-optical techniques in reflection geometry might partially overcome such problem due to their high sensitivity to magnetic and chemical variations across the sample depth. The implementation of a computer code based on a 4X4 matrix formalism describing magneto-optical effects and taking explicitly surface and interface roughness effects into consideration allows obtaining such information by fitting of the experimental spectra. The results reported here show the existence of a surface layer in LCMO thin films with degraded magnetic properties as compared to that of the rest of the film. The thickness of such a layer ranges from 1 to 5 unit cells. This result agrees with theoretical predictions based on the assumption of a localized oxygen deficiency at the surface of the films due to breaking of the crystal symmetry. Moreover the results presented here are also in agreement with the report of an insulating surface layer by Abad et al. [30, 31] and suggest that a charge localization mechanism might act at the surface of LCMO thin films as indicated by theoretical studies [8].

**Acknowledgments:**

We acknowledge financial support from Spanish MCyT (MAT2003-4161 and MAT2006- 13572-C02-01), FEDER program and Generalitat de Catalunya (2005SGR-00509).

**Table caption**

**Table 1.-** Summary of magnetic and structural properties for LCMO thin films.

**Table 2.-** Summary of the spin and orbital magnetic moments in Bohr magnetons obtained by means of XMCD in absorption using the surface sensitive TEY detection channel. The total magnetization defined by $m_{XMCD}=m_{orb}+m_{spin}$ is systematically smaller than the macroscopically averaged value ($m_{SQUID}$) measured by means of SQUID magnetometry.

**Table 3.-** Summary of best fit parameters of AR−XMCD corresponding to the case of a double LCMO layer system. $\chi^2_1$ and $\chi^2_2$ are the final values of the minimization function obtained in the fit cases 1 and 2, respectively (see text for details). Errors are calculated by the Minuit fitting routine [21].

**Table 1**

| Substrate | $T_{ann}$ (°C) | t (nm) | c (nm) | a=b (nm) | $\sigma_{surface}$ (nm) | $T_C$ (K) | $M_S$ (emu/cm$^3$) |
|---|---|---|---|---|---|---|---|
| $SrTiO_3$ | | 6 | ** | ** | 1,2 | 160 | 370 |
| $SrTiO_3$ | | 50 | 0.3806 | 0.3905 | 0,37 | 191 | 550 |
| $SrTiO_3$ | 1000°C | 50 | 0.3824 | 0.3903 | 0,20 | 270 | 570 |
| $LaAlO_3$ | 1000°C | 16 | 0.3908-0.3887[‡] | 0.386 | 2,0 | 270 | 580 |
| $NdGaO_3$ | | 50 | 0.386 | 0.386 | 1,9 | 271 | 446 |
| Bulk LCMO | | | 0.386 | 0.386 | | 280 | 580 |

**Table 2**

| Sample | $m_{orb}$ ($\mu_B$) | $m_{spin}$ ($\mu_B$) | $m_{XMCD}$ ($\mu_B$) | $m_{Squid}$ ($\mu_B$) | Discrepancy |
|---|---|---|---|---|---|
| 6 nm LCMO/STO AG | 0.046 | -1.70 | 1.65 | 2.34 | 29% |
| 50 nm LCMO/STO AG | 0.018 | -1.01 | 0.99 | 3.47 | 71% |
| 50 nm LCMO/STO $T_{ann}$=1000°C | 0.077 | -1.83 | 1.75 | 3.61 | 52% |
| 50 nm LCMO/NGO AG | 0.034 | -1.82 | 1.79 | 2.82 | 36% |
| 16 nm LCMO/LAO $T_{ann}$=1000°C | 0.066 | -2.85 | 2.78 | 3.67 | 24% |
| Bulk LCMO | | | | 3.67 | |

**Table 3**

| Sample<br>Fit variable | LCMO/STO<br>6 nm | LCMO/STO<br>50 nm | LCMO/STO<br>50 nm<br>$T_{ann}$=1000°C | LCMO/LAO<br>16 nm<br>$T_{ann}$=1000°C | LCMO/NGO<br>50 nm |
|---|---|---|---|---|---|
| t (nm) | 4.72 ± 0.08 | 51.37 ± 0.04 | 51.1 ± 0.5 | 16.16 ± 0.07 | 42.41 ± 0.07 |
| $MO_1$ | 1.0 ± 0.1 | 1.0 ± 0.1 | 1.0 ± 0.1 | 1.00 ± 0.04 | 1.0 ± 0.5 |
| t(2) (nm) | 1.06 ± 0.02 | 1.059 ± 0.002 | 1.0 ± 0.2 | 0.50 ± 0.03 | 2.1 ± 0.1 |
| $MO_2$ | 0.00 ± 0.04 | 0.00 ± 0.02 | 0.3 ± 0.1 | 0.56 ± 0.04 | 0.3 ± 0.1 |
| $\sigma_{f/s}$ | 0.61 ± 0.02 | 0.78 ± 0.01 | 1.1 ± 0.2 | 0.9 ± 0.1 | 2.9 ± 0.3 |
| $\sigma_{fSurface}$ | 0.47 ± 0.02 | 0.1 ± 0.2 | 0.16 ± 0.12 | 0.5 ± 0.1 | 2.7 ± 0.4 |
| $\sigma_{1/2}$ | 0 ± 3 | 0.70 ± 0.01 | 0.5 ± 0.2 | 0 ± 3 | 2.9 ± 0.4 |
| $X^2_1$ | 1.191 | 0.228 | 0.136 | 0.049 | 0.2801 |
| $X^2_2$ | 0.216 | 0.145 | 0.069 | 0.045 | 0.088 |

**Figure caption**

**Figure 1.-** (Color online) Simulations of XMCD asymmetry spectra in reflection (AR−XMCD) for various values of the surface roughness (left panel) and of the film/substrate interface roughness (right panel) corresponding to a LCMO/LAO film of t=16 nm. The optical and magneto-optical constants for LCMO are those corresponding to the annealed LCMO/LAO sample. Both panels reveal the sensitivity of magneto-optical techniques in reflection to structural properties.

**Figure 2.-** (Color online) Simulated AR−XMCD spectra for two LCMO layers on STO with thicknesses 6 nm and 50 nm. The optical and magneto-optical constants for LCMO are those corresponding to the annealed LCMO/LAO sample. The surface roughness has been set to 0.4 nm and that of the film/substrate interface to 0.8 nm. Panels a) and b) The strong sensitivity of R-XMCD to non-magnetic layers is demonstrated. The presence of a 1 nm non magnetic surface layer strongly affects the spectral shape.

**Figure 3.-** (Color online) Element specific hysteresis loops obtained by means of XMCD for different photon energies for the t=16 nm annealed LCMO/LAO sample. The magnetic loops have been obtained at various photon energies across the Mn L-edge corresponding to different absorptions (inset), i.e. different depths. No differences are observed.

**Figure 4.-** (Color online) a) c) and e) panels show experimental reflection XMCD asymmetry spectra and fit for the 6 nm and 50 nm as-grown and the 50 nm annealed LCMO/STO films, respectively. Bottom panels b), d) and f ) depicts the comparison between experimental T−MOKE asymmetry and simulations using the fit variables for this same set of samples (see table 3).

**Figure 5.-** (Color online) a) and c) panels show the experimental reflection XMCD asymmetry spectra and fit for the 16 nm annealed LCMO/LAO and 50 nm as-grown LCMO/NGO films, respectively. Bottom panels b) and d) depict the comparison between experimental T−MOKE asymmetry and simulations using fit variables (see table 3) for this same set of samples.

**Figure 1**

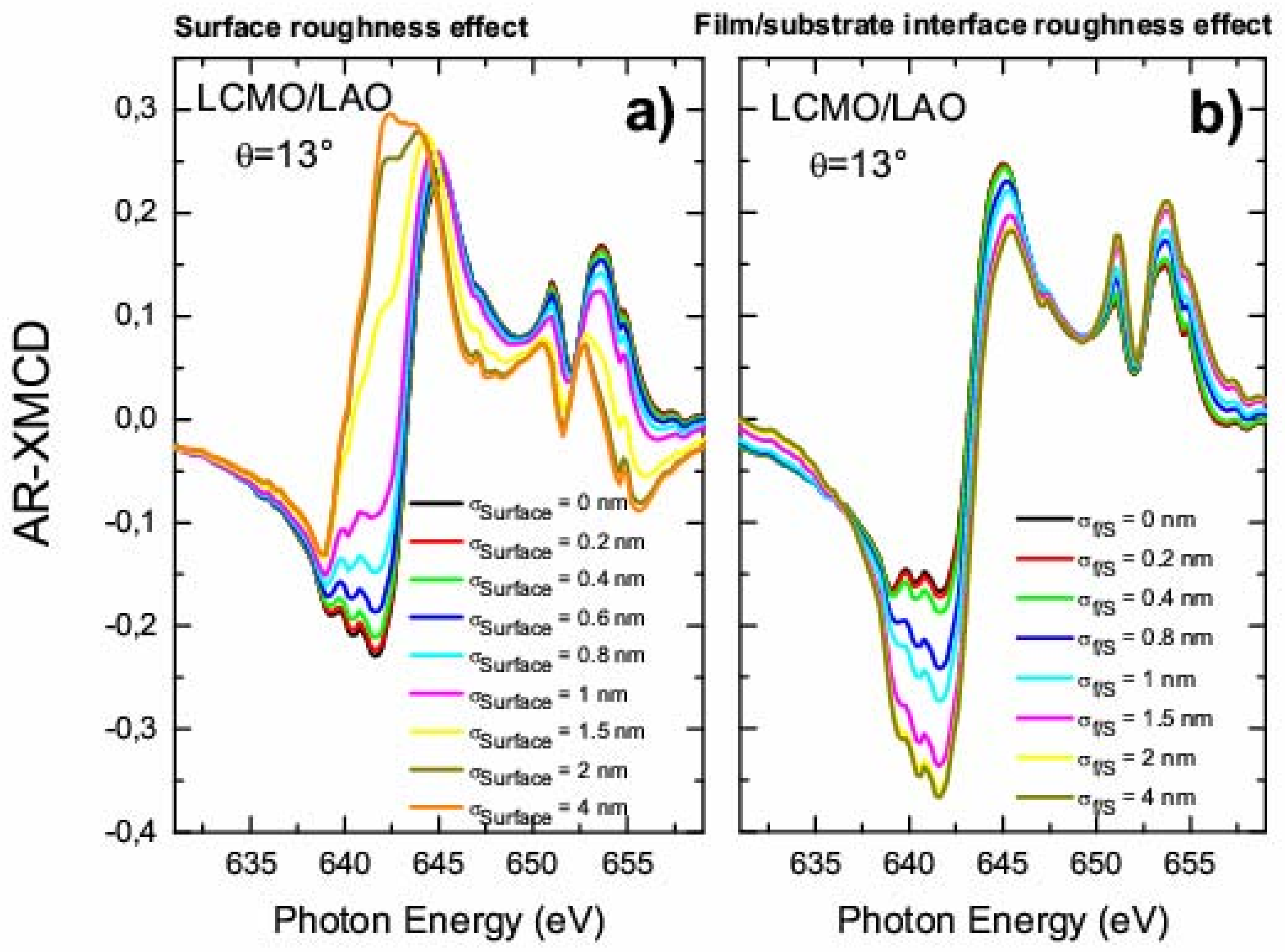
Surface roughness effect
Film/substrate interface roughness effect
LCMO/LAO
θ=13°
a)
LCMO/LAO
θ=13°
b)
AR-XMCD
0,3
0,2
0,1
0,0
-0,1
-0,2
-0,3
-0,4
635
640
645
650
655
Photon Energy (eV)
σSurface = 0 nm
σSurface = 0.2 nm
σSurface = 0.4 nm
σSurface = 0.6 nm
σSurface = 0.8 nm
σSurface = 1 nm
σSurface = 1.5 nm
σSurface = 2 nm
σSurface = 4 nm
σf/S = 0 nm
σf/S = 0.2 nm
σf/S = 0.4 nm
σf/S = 0.8 nm
σf/S = 1 nm
σf/S = 1.5 nm
σf/S = 2 nm
σf/S = 4 nm

**Figure 2**

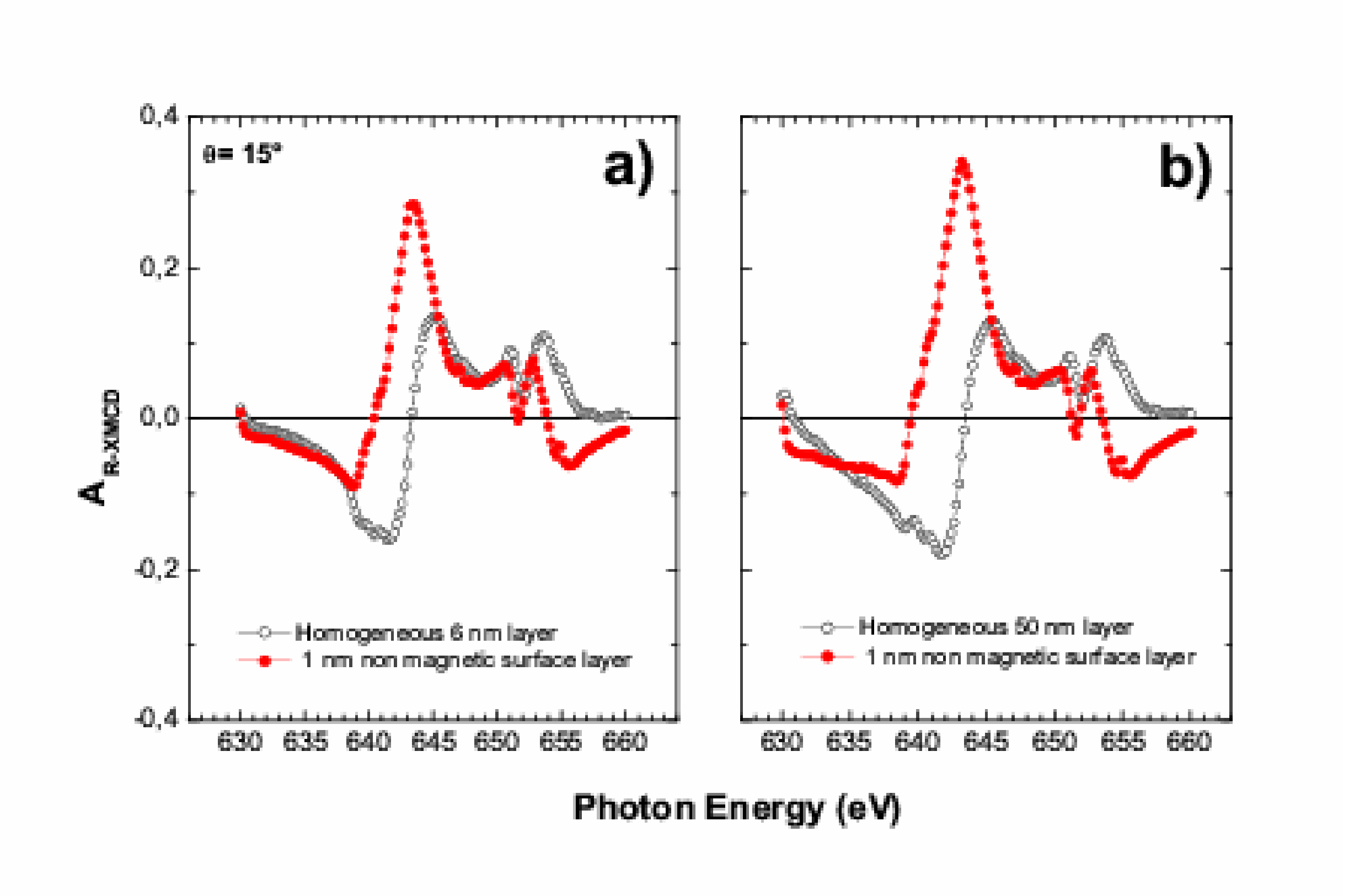

**Figure 3**

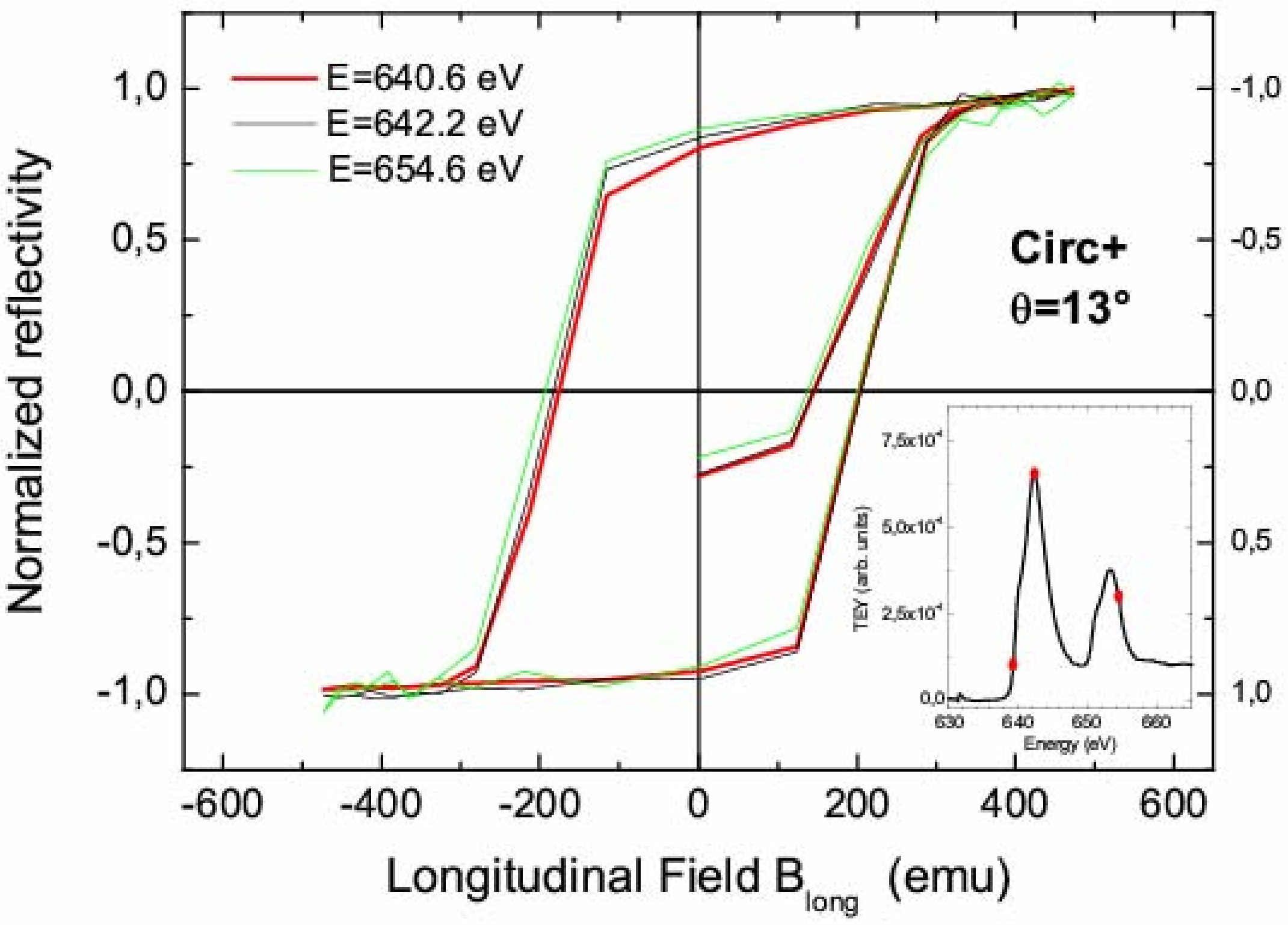

**Figure 4**

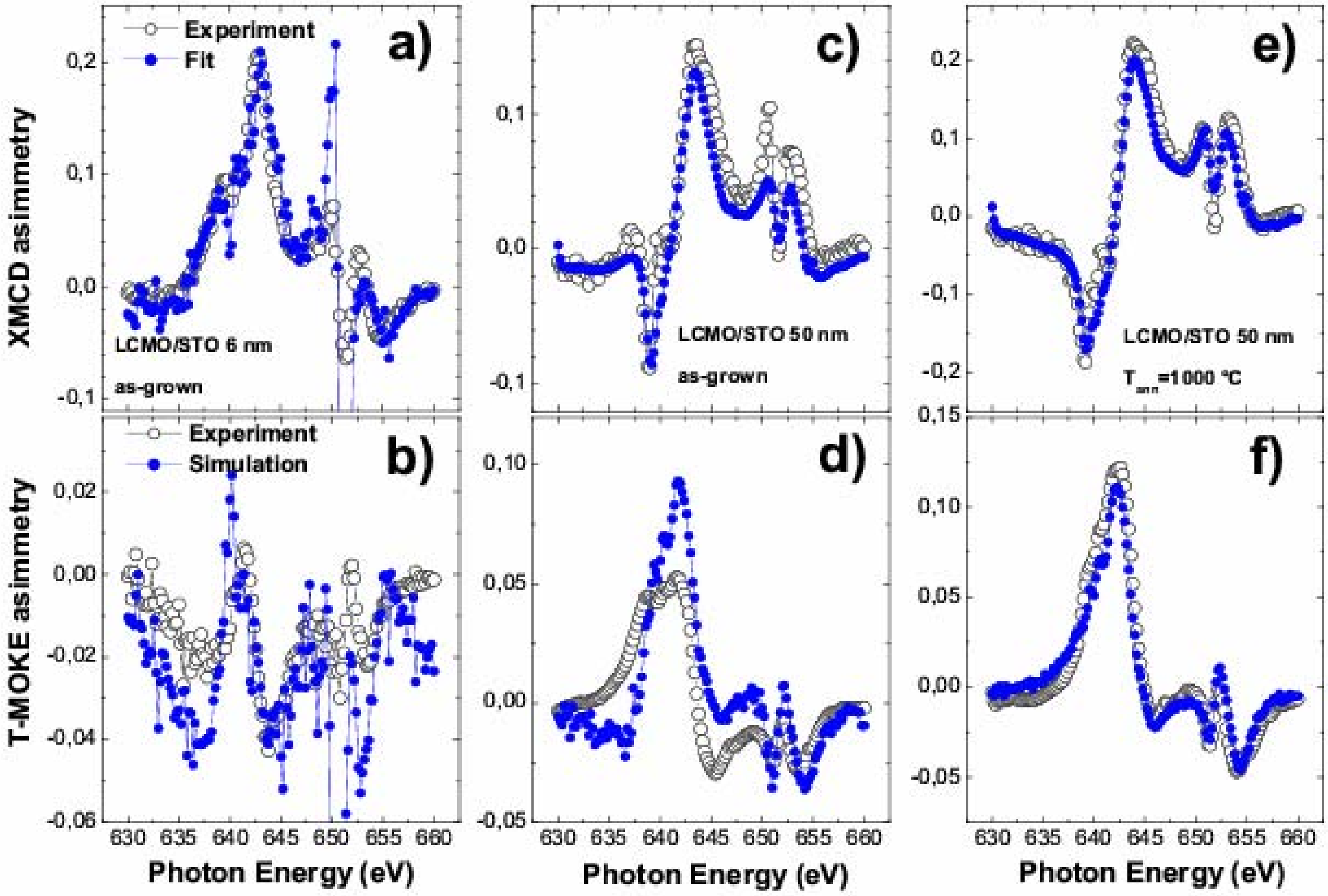

**Figure 5**

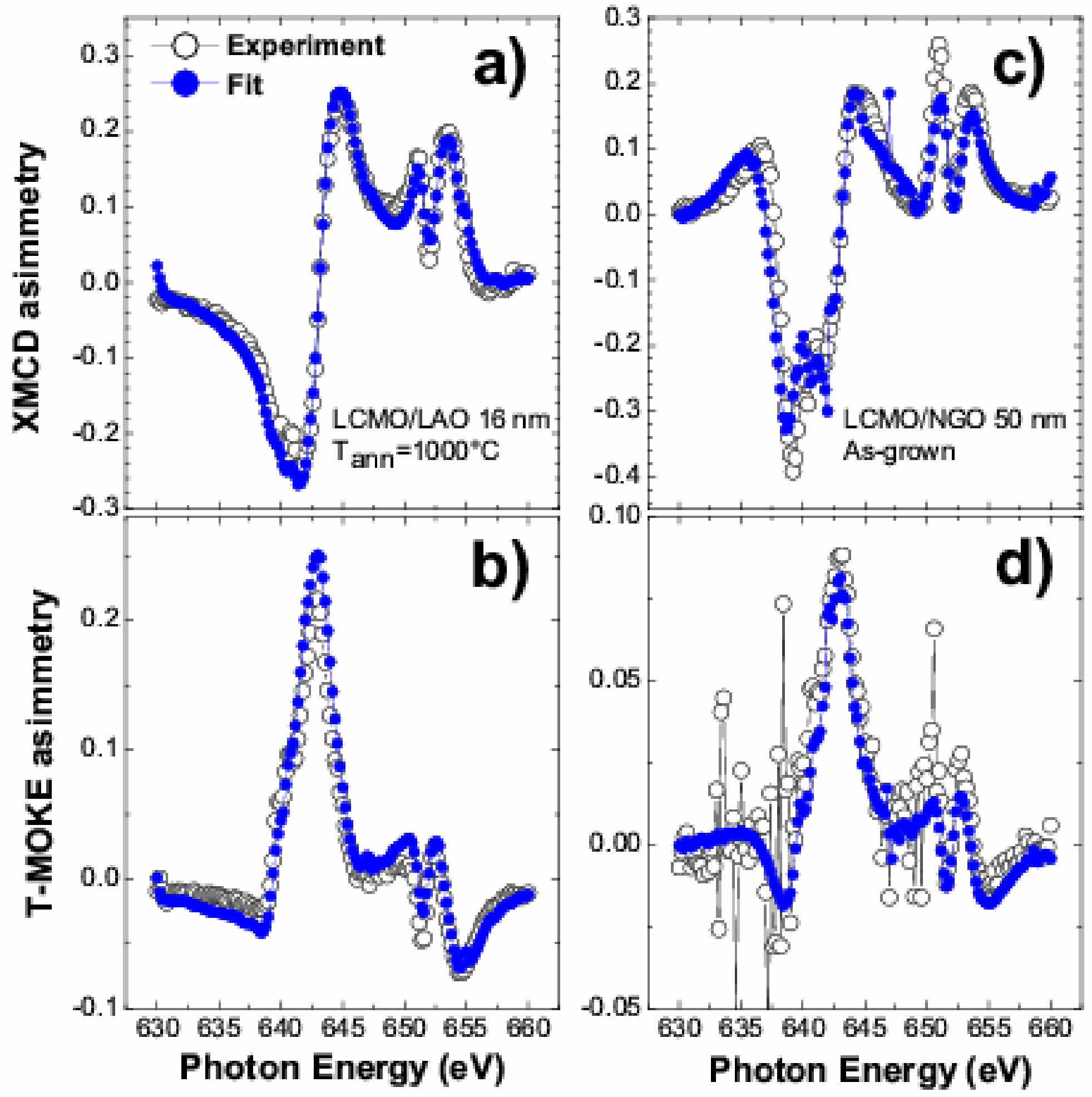